\documentclass{article}
\newlength{\dinwidth}
\newlength{\dinmargin}
\makeatletter
\DeclareSymbolFont{AMSb}{U}{msb}{m}{n}
\DeclareMathAlphabet\frak{U}{euf}{m}{n}
\DeclareSymbolFontAlphabet{\Bbb}{AMSb}
\makeatother
\newcommand{\mb}[1]{\mbox{\boldmath$#1$\unboldmath}}%
\newcommand{\bi}[1]{\mb{#1}}
\newlength{\mylength}
\newcommand{\dz}{\frac{d}{d z}}
\renewcommand{\d}{d}
\newcommand{\wur}[2]{ \left[\begin{array}{ccccccccc} & & & & & & #1 &
                            &  \\  #2  \end{array} \right]}%
\newcommand{\mvek}[2]{\left( #1 | #2 \right)}%
\newcommand{\tach}[1]{\negthinspace \left| \mb{#1} \right\rangle}%
\newcommand{\com}[1]{\big[ #1 \big]}

\newcommand{\Res}{{\rm Res}}
\newcommand{\End}{{\rm End}}
\newcommand{\id}{{\rm id}}
\newcommand{\E}{\mbox{$E_{10}$}}%
\newcommand{\Z}{{\Bbb{Z}}}
\newcommand{\Q}{{\Bbb{Q}}}
\newcommand{\N}{{\Bbb{N}}}
\newcommand{\R}{{\Bbb{R}}}
\newcommand{\C}{{\Bbb{C}}}
\newcommand{\2}{\textstyle\frac{1}{2}}%
\newcommand{\1}{\mb{1}}
\newcommand{\rL}[1]{{\rm L}_{(#1)}}%
\newcommand{\X}{\! \cdot \!}%
\newcommand{\no}{\mbox{\bf \LARGE :}}
\newcommand{\nox}{{}_\times^\times}
\newcommand{\Vir}{{\sf Vir}}
\newcommand{\e}{{\rm e}}
\newcommand{\sC}{{\sf C}}%
\newcommand{\va}{\bi{a}}%
\newcommand{\vk}{\bi{k}}%
\newcommand{\vp}{\bi{p}}%
\newcommand{\vr}{{\bi{r}}}%
\newcommand{\vs}{{\bi{s}}}%
\newcommand{\vv}{{\bi{v}}}%
\newcommand{\vx}{\bi{x}}%
\newcommand{\balpha}{\mb{\alpha}}%
\newcommand{\bdelta}{\mb{\delta}}%
\newcommand{\bomega}{\mb{\omega}}%
\newcommand{\bLambda}{\mb{\Lambda}}%
\newcommand{\bxi}{\mb{\xi}}%
\newcommand{\bri}[1]{{\bi{r}_{#1}}}%
\newcommand{\cF}{{\cal F}}%
\newcommand{\cL}{{\cal L}}%
\newcommand{\cP}{{\cal P}}%
\newcommand{\cV}{{\cal V}}%
\newcommand{\cW}{{\cal W}}%
\newcommand{\fg}{{\frak{g}}}

\newcommand{\fl}{}
\newcommand{\case}[2]{{\textstyle\frac{#1}{#2}}}
\makeatletter
\def\eqalign#1{\null\vcenter{\def\\{\cr}\openup\jot\m@th
  \ialign{\strut$\displaystyle{##}$\hfil&$\displaystyle{{}##}$\hfil
      \crcr#1\crcr}}\,}
\def\cases#1{%
     \left\{\,\vcenter{\def\\{\cr}\normalbaselines\openup1\jot\m@th%
     \ialign{\strut$\displaystyle{##}\hfil$&\tqs
     \rm##\hfil\crcr#1\crcr}}\right.}%
\makeatother
\begin{document}

\thispagestyle{empty}
\renewcommand{\thefootnote}{\fnsymbol{footnote}}
\begin{flushright} hep-th/9610182 \\
                   IASSNS-HEP-96-103 \end{flushright}
\vspace*{2cm}
\begin{center}
{\LARGE \sc Explicit determination of a 727-dimensional \\[5mm]
            root space of the hyperbolic Lie algebra $E_{10}$}\\
 \vspace*{1cm}
       {\sl O. B\"arwald\footnote{Supported by Studienstiftung des
            Deutschen Volkes}}\\
 \vspace*{2mm}
  II. Institut f\"ur Theoretische Physik, Universit\"at Hamburg, \\
  Luruper Chaussee 149, 22761 Hamburg, Germany \\
 \vspace*{4mm}
       {\sl and}\\
 \vspace*{4mm}
       {\sl Reinhold W. Gebert\footnote{Supported by
            Deutsche Forschungsgemeinschaft under Contract No.~{\it Ge
            963/1-1}}} \\
  \vspace*{2mm}
  Institute for Advanced Study, School of Natural Sciences, \\
  Olden Lane, Princeton, NJ 08540, U.S.A. \\
 \vspace*{6mm}
\vspace*{1cm}
\begin{minipage}{11cm}\footnotesize
  The 727-dimensional root space associated with the level-2 root
  $\bLambda_1$ of the hyperbolic Kac--Moody algebra $E_{10}$ is
  determined using a recently developed string theoretic approach to
  hyperbolic algebras. The explicit form of the basis reveals a
  complicated structure with transversal as well as longitudinal
  string states present.
\end{minipage}
\end{center}
\renewcommand{\thefootnote}{\arabic{footnote}}
\setcounter{footnote}{0}
\newpage
\section{Introduction}
Finite and affine Lie algebras represent well-understood classes of
Kac--Moody algebras (see e.g.\ \cite{Kac90,MooPia95}). The Kac--Moody
algebras associated with indefinite Cartan matrices, however, have so
far resisted attempts at systematic and conceptual understanding. The
difficulties arise from the very definition of these Lie algebras in
terms of generators and relations. This definition entails that we
have no basis for indefinite Kac--Moody algebras and so must form
arbitrary multiple commutators of the basic Chevalley generators and
must take into account the Jacobi identity and the complicated Serre
relations. Yet this approach is intractible from a practical point of
view since it would be extremely cumbersome to actually {\it write
down} Lie algebra elements as multiple commutators. In fact, there is
not a single example of an indefinite Kac--Moody algebra for which all
root multiplicities, let alone an explicit basis, are known. Therefore
the crucial first step towards a thorough analysis of these
infinite-dimensional Lie algebras is the search for other realizations
of them (which would be analogous to the realization of affine Lie
algebras as central extensions of loop algebras).

Surprisingly, although perhaps not unexpectedly, Lorentzian Kac--Moody
algebras, which are indefinite Kac--Moody algebras with Lorentzian
signature of the associated root lattice, naturally appear in
theoretical physics, namely, as subalgebras of the Lie algebra of
physical states of a completely compactified chiral bosonic string
with the Lorentzian root lattice as momentum lattice (cf.\
\cite{Fren85,GodOli85,Borc86}). It turns out that for the
uncompactified string a complete basis for the space of physical
states is provided by the so-called DDF construction
\cite{DeDiFu72,Brow72}. So we do have explicit realizations of the
bigger Lie algebra of physical states and the problem is now to single
out those DDF states which lead to a basis of the embedded Lorentzian
Kac--Moody algebra.

In \cite{GebNic95}, a discrete version of the DDF construction was
developed and exploited to tackle the above problem for the case of
the hyperbolic Kac--Moody algebra $E_{10}$. It turned out that the
level-1 sector (which is known to be isomorphic to the basic
representation of the affine subalgebra) is spanned by the transversal
DDF states whereas the longitudinal states are all ``missing'' in the
sense that they lie in the orthogonal complement of the hyperbolic
algebra within the full Lie algebra of physical states. As a
demonstration of the utility of this new approach, the root space of
the fundamental level-2 root $\bLambda_7$ was analyzed and a
192-dimensional DDF basis was found. The latter involved not only
transversal states but also string states with longitudinal
excitations.

In this paper, we apply the DDF analysis to the fundamental level-2
root $\bLambda_1$ and exhibit a 727-dimensional basis for the
associated root space of the hyperbolic algebra $E_{10}$. From the
complicated structure in terms of transversal and longitudinal
polarization tensors (see page \pageref{Basis}), we draw the following
conclusions. First, it is clear that the longitudinal DDF operators
(which in general form a Virasoro algebra with central charge $26-d$)
will play an essential role in understanding the higher-level sectors
of the hyperbolic algebra (cf.\ \cite{GeKoNi96,GebNic97}). Second,
hyperbolic Kac--Moody algebras seem to have some additional hidden
structure which cannot be revealed by the perturbative string model
alone (see e.g.\ \cite{GebNic94} for further speculations).

Let us briefly describe how the paper is organized. After introducing
the string model in the framework of vertex algebras, we review the
discrete DDF construction for the example $E_{10}$. Finally, we
present the explicit basis for the analyzed root space. The necessary
calculations are collected in an appendix. For further details we
refer the reader to the diploma thesis \cite{Baer96}, on which the
present paper is based.
\section{Vertex algebras and compactified strings}
We shall study one chiral sector of a closed bosonic string moving on
a Minkowski torus as spacetime, i.e., with all target space
coordinates compactified.  Uniqueness of the quantum mechanical wave
function then forces the center of mass momenta of the string to form
a lattice with Minkowskian signature.

Upon ``old'' covariant quantization this system turns out to realize a
mathematical structure called vertex algebra \cite{Borc86}. For a
detailed account of this topic the reader may wish to consult the
review \cite{Gebe93}. Here, we will follow closely \cite{GebNic95},
omitting most of the technical details.

We work with {\bf formal variables} $z_1,z_2,\ldots$ and formal
Laurent series, an algebraic approach to complex analysis. For a
vector space $S$, we define the $\C$-vector space of formal series as
$S[\![z,z^{-1}]\!]= \{\sum_{n\in\Z}s_n z^n| s_n \in S\}$. For a formal
series we use the residue notation
\begin{equation}
\Res_z\bigg[ \sum_{n \in \Z} s_n z^n\bigg] = s_{-1}.
\end{equation}
A thorough introduction to formal calculus can be found in
\cite{FLM88}. In the string model we shall consider the formal
variables as having their origin as complex worldsheet coordinates.

A {\bf vertex algebra} $(\cF,\cV,\1,\bomega)$ is a $\Z$-graded vector
space $\cF= \bigoplus_{n\in\Z}\cF^n$, equipped with an injective
linear map $\cV:\cF \mapsto (\End \cF)[\![z,z^{-1}]\!]$, which assigns
to every {\bf state} $\psi \in \cF$ a {\bf vertex operator}
$\cV(\psi,z)$.  As operator-valued formal Laurent series, vertex
operators are completely determined by their mode operators defined by
\begin{equation}
\cV(\psi,z)=\sum_{n \in \Z} \psi_nz^{-n-1},
\end{equation}
where $\psi_n\in\End \cF$ for all $n$. The state space $\cF$ of a
vertex algebra contains two preferred elements: the {\bf vacuum} $\1$
and the {\bf conformal vector} $\bomega$ whose associated vertex
operators are given by the identity $\id_\cF$ and the stress--energy
tensor
\begin{equation}
\cV(\bomega,z)=\sum_{n \in \Z} \rL{n}z^{-n-2},
\end{equation}
respectively. The latter provides the generators $\rL{n}$ of the {\bf
  Virasoro algebra} \Vir{} such that the grading of $\cF$ is obtained
by the eigenvalues of $\rL{0}$ and the role of a translation generator
is played by $\rL{-1}$ satisfying $\cV(\rL{-1}\psi)=\dz \cV(\psi,z)$.
Finally, there is a crucial identity relating products and
iterates of vertex operators called the {\bf (Cauchy--)Jacobi
  identity}:
\begin{equation}
\label{eq:jacobi}
\sum_{i\ge0}(-1)^i{l \choose i}
 (\psi_{l+m-i}\phi_{n+i}-(-1)^l\phi_{l+n-i}\psi_{m+i})
=\sum_{i\ge0}{m \choose i}(\psi_{l+i}\phi)_{m+n-i},
\end{equation}
for all $\psi,\phi\in\cF,\ l,m,n\in\Z$.

An important property of vertex algebras is the {\bf skew symmetry}
\begin{equation}
\label{eq:skewsym}
\cV(\psi,z)\varphi = \e^{z\rL{-1}}\cV(\varphi,-z)\psi,
\end{equation}
which shows that the vertex operator $\cV(\psi,z)$ ``creates'' the
state $\psi \in \cF$ from the vacuum, viz
\begin{equation}
\cV(\psi,z)\1=\e^{z\rL{-1}}\psi.
\end{equation}
In string theory a special role is played by the subspace $\cP^1 \subset
\cF^1$, the space of {\bf (conformal) highest weight vectors} or {\bf
  primary states} of weight 1, satisfying
\begin{eqnarray}
\label{eq:primstatesa}
\rL{0}\psi=\psi \\
\label{eq:primstatesb}
\rL{n}\psi=0 \qquad \forall n >0.
\end{eqnarray}
We shall call the states in $\cP^1$ {\bf physical states} from now on.
The vertex operators associated with physical states enjoy rather
simple commutation relations with the generators of \Vir{}; in terms
of the mode operators we have
\begin{equation}
\label{eq:primmodes}
\com{\rL{n},\psi_m}=-m\psi_{m+n},
\end{equation}
so that in particular the zero modes of physical states commute with
the Virasoro constraints.

The quotient space $\cF / \rL{-1} \cF$ carries the structure of a Lie
algebra with bracket defined by \cite{Borc86}
\begin{equation}
\com{\psi,\varphi}:=\Res_z\left[\cV(\psi,z)\varphi\right]=\psi_0\varphi,
\end{equation}
where the antisymmetry and the Jacobi identity follow from the skew
symmetry (\ref{eq:skewsym}) and the Cauchy--Jacobi identity
(\ref{eq:jacobi}), respectively.
The Lie algebra $\cF / \rL{-1} \cF$ is too large for a further
analysis. We will therefore concentrate on a subalgebra which is also
relevant for string theory, namely the {\bf Lie algebra of physical
  states}:
\begin{equation}
\fg_\cF := \cP^1 / \rL{-1} \cP^0.
\end{equation}

In general, the physics described by vertex algebras corresponds to
meromorphic bosonic two-dimensional conformal quantum field theories
(see e.g.\ \cite{Godd89b}). Here, we will consider the vertex algebra
associated with one chiral sector of a first quantized bosonic string
theory moving on a Minkowski torus as target space. We will briefly
review the main ingredients of the model, the details of which can be
found in \cite{GebNic95}.

Let $\Lambda$ be an even nongenerate lattice of rank $d< \infty$,
representing the lattice of allowed center-of-mass momenta for the
string. To each lattice point we assign a {\bf zero mode state}
$\e^{\vr}$, which we alternatively denote by $\tach{r}$ and which
provides a highest weight vector for a $d$-fold Heisenberg algebra
$\hat{\mbox{\bf h}}$ of oscillators $\alpha_m^\mu, n \in \Z, 1 \le \mu
\le d$,
\begin{equation}
\alpha^\mu_0\tach{r}=r^\mu\tach{r},\qquad
\alpha^\mu_m\tach{r}=0\quad \forall n >0,
\end{equation}
where
\begin{equation}
[\alpha^\mu_m,\alpha^\nu_n]=m\eta^{\mu\nu}\delta_{m+n,0}.
\end{equation}
These zero mode states realize the $\epsilon$-twisted group algebra
$\R\{\Lambda\}$ for some 2-cocycle $\epsilon$. For notational
convenience we put $\vr(m)\equiv\vr \X \balpha_m$ and
$\vr(z)\equiv\sum_{m \in \Z} \vr(m) z^{-m-1}$. The Fock space is
obtained by collecting the Heisenberg modules built on all zero mode
states, viz
\begin{equation}
\cF := S\big(\hat{\mbox{\bf h}}^-\big) \otimes \R\{\Lambda\}.
\end{equation}
We now assign to each state $\psi\in \cF$ a {\bf vertex operator}
$\cV(\psi,z)$.
For a zero mode state $\e^\vr$ we define
\begin{equation}
\label{eq:vertexop}
\cV(\e^{\vr},z) := \textstyle\exp(\int \vr_<(z) \d z) \e^{\vr}
z^{\vr(0)} \exp(\int \vr_>(z) \d z),
\end{equation}
with $\vr_<(z)=\sum_{m \in \N} \vr(-m) z^{m-1}$ and $\vr_>(z)=\sum_{m
\in \N} \vr(m) z^{-m-1}$. And for $\psi = \prod_{j=1}^N
\vs_j(-n_j)\otimes \e^\vr$, a general homogeneous element of $\cF$, we
have
\begin{equation}
\cV(\psi,z) := \no \cV(\e^\vr,z) \prod_{j=1}^N \frac{1}{(n_j-1)!}
\bigg( \dz \bigg)^{n_j-1} \vs_j(z) \no.
\end{equation}
This definition can be extended by linearity to the whole of
$\cF$. These operators indeed fulfill the axioms of a vertex algebra
\cite{FLM88}.

A special role is played by the lattice vectors of length 2 which are
called the {\bf real roots} of the lattice. We associate with every real
root a {\bf reflection} by $w_\vr(\vx) = \vx - (\vx \,\X\, \vr) \vr$
for $\vx \in \Lambda$. The hyperplanes perpendicular to these divide
the real vector space $\R \otimes_\Z \Lambda$ into regions called {\bf
  Weyl chambers}. The reflections in the real roots of $\Lambda$
generate a group called the {\bf Weyl group} $\cW$ of $\Lambda$, which
acts simply transitively on the Weyl chambers. Fixing one chamber to
be the {\bf fundamental Weyl chamber} $\sC$ once and for all, we call
the real roots perpendicular to the faces of $\sC$ and with inner
product at most 0 with elements of $\sC$, the {\bf simple roots}
$\vr_i$ of \sC{}.

The physical states
\begin{equation}
e_i:=\e^{\vr_i},\qquad
f_i:=-\e^{-\vr_i},\qquad
h_i:=\vr_i(-1)\tach{0},
\end{equation}
for all $i$ then obey the following commutation relations (see
\cite{Borc86}):
\begin{eqnarray}
   [h_i,h_j]=0&,& \\ {}
   [h_i,e_j]=A_{ij}e_j&,&\quad [h_i,f_j]=-A_{ij}f_j, \\  {}
   [e_i,f_j]=\delta_{ij}h_i&,& \\ {}
   ({\rm ad}\,e_i)^{1-A_{ij}}e_j=0&,&\quad
   ({\rm ad}\,f_i)^{1-A_{ij}}f_j=0\quad\mbox{($i\neq j$)}
\end{eqnarray}
i.e., they generate via multiple commutators the {\bf Kac--Moody
  algebra} $\fg(A)$ associated with the Cartan matrix $A= (A_{ij})=
(\vr_i \X \vr_j)$, which is a subalgebra of the Lie algebra of
physical states $\fg_\Lambda$. In the Euclidean case both algebras
coincide but in general we have a proper inclusion
\begin{equation}
\fg(A) \subset \fg_\Lambda.
\end{equation}
The main problem in this string realization of the hyperbolic Lie
algebra is to determine the elements of $\fg_\Lambda$ \em not \em
contained in $\fg(A)$, which we call {\bf missing} or {\bf
  decoupled states}.

\section{$E_{10}$ and the DDF construction}

Most of the information about the hyperbolic Kac--Moody algebra
$E_{10}$ available in the mathematical literature can be found in
\cite{KaMoWa88}; readers interested in more general information about
infinite dimensional Lie algebras should consult the textbooks
\cite{Kac90} or \cite{MooPia95}. The hyperbolic Lie algebra \E{} is
defined via its Coxeter--Dynkin diagram and the Serre relations
following from it. The root lattice $Q(\E)$ coincides with the unique
10-dimensional even unimodular Lorentzian lattice $I\!I_{9,1}$.  The
latter can be defined as the lattice of all points
$\vx=(x_1,\ldots,x_{9}|x_0)$ for which the $x_m$'s are all in $\Z$ or
all in $\Z+\frac{1}{2}$ and which have integer inner product with the
vector $\mb{l}=(\frac{1}{2},\ldots,\frac{1}{2}|\frac{1}{2})$, all
norms and inner products being evaluated in the Minkowskian metric
$\vx^2=x_1^2+\ldots+x_9^2-x_0^2$.  A basis of simple roots for this
lattice is given by the 10 lattice vectors $$
\begin{array}{l@{\quad=\quad(}r@,r@,r@,r@,r@,r@,r@,r@,r@{|}l}
  \bri{-1} & 0 & 0 & 0 & 0 & 0 & 0 & 0 & 1 & -1 & 0),  \\
  \bri{ 0} & 0 & 0 & 0 & 0 & 0 & 0 & 1 & -1 & 0 & 0),  \\
  \bri{ 1} & 0 & 0 & 0 & 0 & 0 & 1 & -1 & 0 & 0 & 0),  \\
  \bri{ 2} & 0 & 0 & 0 & 0 & 1 & -1 & 0 & 0 & 0 & 0),  \\
  \bri{ 3} & 0 & 0 & 0 & 1 & -1 & 0 & 0 & 0 & 0 & 0),  \\
  \bri{ 4} & 0 & 0 & 1 & -1 & 0 & 0 & 0 & 0 & 0 & 0),  \\
  \bri{ 5} & 0 & 1 & -1 & 0 & 0 & 0 & 0 & 0 & 0 & 0),  \\
  \bri{ 6} & -1 & -1 & 0 & 0 & 0 & 0 & 0 & 0 & 0 & 0),  \\
  \bri{ 7} & \2 & \2 & \2 & \2 & \2 & \2 & \2 & \2 & \2 & \2),  \\
  \bri{ 8} & 1 & -1 & 0 & 0 & 0 & 0 & 0 & 0 & 0 & 0).
\end{array} $$
These simple roots indeed generate the reflection group of $I\!I_{9,1}$.
The corresponding Coxeter--Dynkin diagram looks as follows:
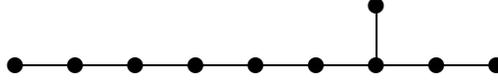
\[ \unitlength1mm
   \begin{picture}(66,10)
   \multiput(1,0)(8,0){9}{\circle*{2}}
   \put(49,8){\circle*{2}}
   \multiput(2,0)(8,0){8}{\line(1,0){6}}
   \put(49,1){\line(0,1){6}}
\end{picture}
\]
The Cartan matrix is $A_{ij}=\vr_i \X \vr_j$. The fundamental Weyl
chamber $\sC$ of $E_{10}$ is the convex cone generated by the {\bf
  fundamental weights $\bLambda_i$} which obey $\bLambda_i \X \vr_j =
- \delta_{ij}$ and are explicitly given by
\begin{equation}
\bLambda_i = - \sum_{j=-1}^8 (A^{-1})_{ij} \vr_j \quad\mbox{for } i=
-1,0,1,\ldots 8
\end{equation}
in terms of the inverse Cartan matrix. In the following, the $E_9$
null root $\bdelta$, dual to the overextended simple root $\vr_{-1}$,
will play an important role; it is
\begin{equation}
\label{eq:nullroot}
\bdelta=\sum_{i=0}^8 n_i
\vr_i=\wur{3}{0&1&2&3&4&5&6&4&2}=\mvek{0,0,0,0,0,0,0,0,1}{1}.
\end{equation}
A useful notion in the theory of hyperbolic Kac--Moody algebras is the
{\bf level} $\ell$ of a given root $\bLambda$, which is defined as the
number of times the simple root $\vr_{-1}$ occurs in it. A useful
formula for $\ell$ is
\begin{equation}
\ell \equiv \ell(\bLambda) := -\bdelta \X \bLambda
\end{equation}
where $\bdelta$ denotes the null root given by (\ref{eq:nullroot}).
Obviously $\ell$ is only invariant under the affine Weyl subgroup
$\cW(E_9)$.

To get a grip on the Lie algebra of physical states $\fg_\Lambda$
associated with the lattice $I\!I_{9,1}$ we now employ the DDF
construction well known from string theory (see e.g.\
\cite{GSW88}). It provides a convenient way of obtaining all physical
states by applying the DDF operators to the tachyonic ground states. A
special feature of subcritical algebras, i.e.\ algebras with rank
$d<26$, is the relevance of {\em longitudinal} DDF operators for the
spectrum of physical states.

Let us recall the basic features of the DDF construction. One starts
with a tachyonic ground state of momentum $\vv$ with $\vv^2=2$ and an
associated light-like vector $\vk(\vv)$ with $\vk^2=0$ and $\vv \X
\vk=1$.  For continuous momenta, such vectors can always be found and
rotated into a convenient frame, but on the lattice this is in general
not possible.  For this reason we have to work with the
rationalization of the root lattice, that is $\Q \otimes_\Z \Lambda$,
to incorporate the intermediate points. To see this let us define the
{\bf DDF decomposition} of a given level-$\ell$ root $\bLambda$ in the
fundamental Weyl chamber by
\begin{equation}
\bLambda=\vv-n\vk
\end{equation}
where $\vk \equiv\vk(\vv) := -\case1\ell\bdelta$, and $n :=
1-\case12\bLambda^2$ is the number of steps required to reach the root
by starting from $\vv$ and decreasing the momentum by $\vk$ at each
step. Notice that we have a factor of $\case1\ell$ in the definition
of $\vk$ at level $\ell$ and so beyond level 1 neither $\vk$ nor $\vv$
will be on the lattice in general.

In addition, we need a set of transversal polarization vectors
$\bxi_i=\bxi_i(\vv,\vk)$ subject to $\bxi_i \X \vv=\bxi_i \X \vk=0$,
which for convenience we choose to be orthonormalized. We use the
letters $i,j,\ldots = 1,\ldots,8$ to label the transversal indices.
Given these data, we can define the {\bf transversal DDF operators}
\cite{DeDiFu72} by
\begin{equation}
A_m^i(\vv) := \Res_z \left[ \bxi_i(z) \cV(\e^{m\vk},z)\right]
\end{equation}
which describe the emission of a photon with momentum $m\vk$
and polarization $\bxi_i$ from the string. Note that no normal-ordering
is required here due to the fact that $\vk$ is light-like and is
orthogonal to the polarization vectors.  Normal-ordering is however
required in the definition of the {\bf longitudinal DDF operators}
\cite{Brow72} given by
\begin{equation}
A_m^- (\vv) := \cL_m(\vv) - \case12\sum_{i=1}^8\sum_{n \in \Z}
               \nox A_n^i(\vv) A_{m-n}^i(\vv)\nox
               +2\delta_{m0}\vk\X\vp
\end{equation}
with
\begin{equation}
\cL_m(\vv) := \Res_z\left[-\cV(\vv(-1)\e^{m\vk},z)
                 +\frac{m}{2} (\vv \X \vk )\dz \log
                 \left(\frac{\vk(z)}{\vv \X \vk}\right)
                 \cV(\e^{m\vk},z)\right]
\end{equation}
where we used $\vk \X \vv=1$. The normal-ordering $\nox \ldots \nox$
places DDF operators with positive index to the left and operators
with negative index to the right. The longitudinal operators $A_m^-$
realize a Virasoro algebra with central charge $c=26-d$, the
transversal operators realize a $(d-2)$-dimensional Heisenberg
algebra, and both families of operators commute with each other. In
the sequel we will be careful to indicate the dependence of the
DDF operators on their tachyonic momenta since we will have to
calculate commutators of DDF operators with \em different \em momenta.

Let us return to the hyperbolic algebra $E_{10}$. Any level-2 root in
\sC{} must be of the form $\bLambda_1+ n \bdelta$ or
$\bLambda_7+n\bdelta$ or $2\bLambda_0+n \bdelta$ for some $n \in \N$.
In \cite{GebNic95} the 192-dimensional root space of the root
$\bLambda_7$ was determined. We will now discuss the next nontrivial
example, the root $\bLambda_1$, dual to the simple root
$\vr_1$. Explicitly, $\bLambda_1$ is given by
\begin{equation}
\bLambda_1= \wur{9}{2 & 4 & 6 & 9 & 12 & 15 & 18 & 12 & 6}
             = \mvek{0,0,0,0,0,0,1,1,1}{3},
\end{equation}
hence $\bLambda_1^2=-6$. The dimension of the associated root space is
$727$. This deviates by one from the number of transversal states,
which is 726. Since, on the other hand, it is well known that the
level-1 states are precisely given by the transversal DDF states, one
might therefore be tempted to conjecture that the 727-dimensional root
space is spanned by the 726 transversal states and one distinguished
longitudinal state. As we will see, however, this is not true and
hints at some additional hidden structure inside $E_{10}$.

In general, one has the level-2 multiplicity formula \cite{KaMoWa88}
\begin{equation}
{\rm mult}(\bLambda)=\xi(3-\case12 \bLambda^2),
\end{equation}
where $\xi$ is defined in terms of the generating
function $ \sum_{n \geq 0} \xi (n) q^n =\phi (q)^{-8} [ 1- \phi
  (q^2)/\phi (q^4) ]$, $\phi (q)=\prod_{l=1}^\infty
(1-q^l)$ denoting the Euler function.  Recall that we noticed earlier
that the appearance of longitudinal states is generic for {\em any}
level-2 root, and that therefore this deviation comes as no surprise,
but from this example the emergence of the longitudinal states could
have been anticipated. Our general DDF decomposition simplifies to
\begin{equation}
\label{DDF_zerlegung}
\bLambda_1 = \vr + \vs + m \bdelta
\end{equation}
where $\vr$ and $\vs$ are two real positive level-1 roots. In general
there will be many different ways to split $\bLambda_1$ in this
manner. These are all related by elements of the finite group
$\cW(\bLambda_1)$, the stabilizer of $\bLambda_1$, whereas the
decompositions for fixed $m$ are invariant under the little group
$\cW(\bLambda_1,\bdelta)$, which is the (finite) subgroup of the full
hyperbolic Weyl group leaving both $\bLambda_1$ and $\bdelta$
invariant. Thus, for every $m$, we will choose one of these
decompositions, work out all commutators, and act on the resultant
equations with the little Weyl group to obtain all possible
states. For different values of $m$ the situation is not that simple,
due to the fact that the little Weyl group is infinite dimensional in
this case. Therefore we will deal with these decompositions case by
case and combine the resulting states in the end.

We will need two different decompositions:
\begin{enumerate}
\item $\bLambda_1=\vr+\vs+3 \bdelta$ with
\begin{equation}
\label{eq:decomp}
\eqalign{
\vr:=\wur{0}{1 & 0 & 0 & 0 & 0 & 0 & 0 & 0 & 0}
    =  \mvek{0,0,0,0,0,0,0,1,-1}{0}\\
\vs:=\wur{0}{1 & 1 & 0 & 0 & 0 & 0 & 0 & 0 & 0}
    =  \mvek{0,0,0,0,0,0,1,0,-1}{0}
}
\end{equation}
\item $\bLambda_1=\vr'+\vs'+2 \bdelta$ with
\begin{equation}
\eqalign{
\vr':=\wur{0}{1 & 1 & 1 & 0 & 0 & 0 & 0 & 0 & 0}
    =  \mvek{0,0,0,0,0,1,0,0,-1}{0}\\
\vs':=\wur{3}{1 & 1 & 1 & 3 & 4 & 5 & 6 & 4 & 2}
    =  \mvek{0,0,0,0,0,-1,1,1,0}{1}
}
\end{equation}
\end{enumerate}
Although we will need several sets of polarization vectors adjusted
to the different DDF decompositions, we will present the basis using
the following set, which is adjusted to the first decomposition
(\ref{eq:decomp}):
\begin{equation}
\eqalign{
\bxi_\alpha &\equiv \bxi_\alpha(\vr) = \bxi_\alpha(\vs)
=\bxi_\alpha(\va)\qquad \mbox{for } \alpha = 1, \ldots , 7, \\
\bxi_1 & =\mvek{1,0,0,0,0,0,0,0,0}{0} \\
\vdots \\
\bxi_6 & =\mvek{0,0,0,0,0,1,0,0,0}{0}\\
\bxi_7 & =\case12 \sqrt{2} \mvek{0,0,0,0,0,0,1,1,1}{1}  \\
\bxi_8(\va) & = \case12 \sqrt{2} \mvek{0,0,0,0,0,0,1,-1,0}{0} \\
\bxi_8(\vr) &= \case12 \sqrt{2} \mvek{0,0,0,0,0,0,-1,1,1}{1} \\
\bxi_8(\vs) &= \case12 \sqrt{2} \mvek{0,0,0,0,0,0,1,-1,1}{1}.
}
\end{equation}
For the little group we have $\cW(\bLambda_1,\bdelta)\equiv \Z_2 \times
\cW(\rm{E}_7)$, the group generated by the 8 fundamental
reflections $\{w_0,w_2,w_3,w_4,w_5,w_6,w_7,w_8 \}$.  On the above system of
polarization vectors the group acts as follows:
\begin{equation}
\eqalign{
w_0(\bxi_8)=-\bxi_8,\\
w_2(\bxi_5)=\bxi_6  , w_2(\bxi_6)=\bxi_5,\\
w_3(\bxi_4)=\bxi_5  , w_3(\bxi_5)=\bxi_4, \\
w_4(\bxi_3)=\bxi_4 ,  w_4(\bxi_4)=\bxi_3,\\
w_5(\bxi_2)=\bxi_3 ,  w_5(\bxi_3)=\bxi_2,\\
w_8(\bxi_1)=\bxi_2 ,  w_8(\bxi_2)=\bxi_1,\\
w_6(\bxi_1)=-\bxi_2 ,  w_6(\bxi_2)=-\bxi_1,\\
\textstyle w_7(\bxi_i)=\bxi_i-\frac{1}{4}\sum_{j=1}^6
\bxi_j-\frac{1}{4}\sqrt{2}\bxi_7 \quad i=1 \ldots 6, \\
\textstyle w_7(\bxi_7)=\case12\bxi_7 -\frac{1}{4}\sqrt{2} \sum_{j=1}^6
\bxi_j. }
\end{equation}
The DDF construction provides us with an explicit basis for the
physical states of momentum $\bLambda_1$. We have 726 transversal and
54 longitudinal states,
\begin{equation}
\begin{array}{r}
A_{-1}^i(\mb{a})A_{-1}^j(\mb{a})A_{-1}^k(\mb{a})A_{-1}^l(\mb{a})\tach{a},\\
A_{-1}^i(\mb{a})A_{-1}^j(\mb{a})A_{-2}^k(\mb{a})\tach{a},\\
A_{-2}^i(\mb{a})A_{-2}^j(\mb{a})\tach{a},\\
A_{-1}^i(\mb{a})A_{-3}^j(\mb{a})\tach{a},\\
A_{-4}^i(\mb{a})\tach{a}  ,\\
A_{-1}^i(\mb{a})A_{-1}^j(\mb{a})A_{-2}^-(\mb{a})\tach{a}  ,\\
A_{-2}^i(\mb{a})A_{-2}^-(\mb{a})\tach{a}  ,\\
A_{-1}^i(\mb{a})A_{-3}^-(\mb{a})\tach{a}  ,\\
A_{-2}^-(\mb{a})A_{-2}^-(\mb{a})\tach{a}  ,\\
A_{-4}^-(\mb{a})\tach{a}.
\end{array}
\end{equation}
Therefore we can express any element of the root space
$E_{10}^{(\bLambda_1)}$ as a linear combination of these states.
Since the $E_{10}$ states are precisely the elements of
$\fg_{I\!I_{9,1}}^{(\bLambda_1)}$ that are expressible by commutators
of level-1 elements, the problem is to work out the ``Clebsch--Gordan
coefficients'' occurring in the expansion
\begin{eqnarray*}
\lefteqn{
\com{ A^{i_1}_{-m_1}(\vs)\ldots A^{i_M}_{-m_M}(\vs)\tach{s},
A^{j_1}_{-n_1}(\vr)\ldots A^{j_N}_{-n_N}(\vr)\tach{r}}=} \\
& &\sum_{p_1+\ldots+q_Q=n \atop k_1,\ldots,k_P}
 c_{k_1\ldots k_P}^{i_1\ldots i_M j_1\ldots j_N}
   A^{k_1}_{-p_1}(\va) \ldots A^{k_P}_{-p_P}(\va)
   A_{-q_1}^-(\va)\ldots A_{-q_Q}^-(\va) \tach{a},
\end{eqnarray*}
into which all the information on how $E_{10}$ sits inside the Lie
algebra of physical states is encoded. We found that the following 727 states
form a complete basis of the root space $E_{10}^{(\bLambda_1)}$:
\[\fl \label{Basis}
\begin{array}{r}
{A_{-4}^i \tach{a}} ,\\
{A_{-2}^i A_{-2}^j \tach{a}} ,\\
{A_{-2}^i A_{-2}^- \tach{a}} ,\\
{A_{-2}^- A_{-2}^-  \tach{a}} ,\\
{A_{-1}^\alpha A_{-1}^\beta A_{-1}^\gamma A_{-1}^7 \tach{a}} ,\\
{\{A_{-1}^\mu A_{-3}^\nu - A_{-3}^\mu A_{-1}^\nu \}\tach{a}} ,\\
{\{A_{-1}^8 A_{-3}^\mu + 3 A_{-1}^\mu A_{-3}^8  \}\tach{a}} ,\\
\settowidth{\mylength}{${}^{(1)}$}
{\{A_{-1}^i A_{-1}^i A_{-1}^i A_{-1}^i - 2 A_{-1}^i
A_{-3}^i\}\tach{a}}{}^{(1)}\hspace*{-\mylength},\\
{\{A_{-1}^\alpha A_{-1}^\alpha A_{-1}^\alpha A_{-1}^\beta - A_{-3
}^\alpha A_{-1}^\beta \}\tach{a}} ,\\
{\{A_{-1}^\alpha A_{-1}^\alpha A_{-1}^\alpha A_{-1}^7 + 5  A_{-3
}^\alpha A_{-1}^7 \}\tach{a}} ,\\
{\{A_{-1}^\alpha A_{-1}^7 A_{-1}^7 A_{-1}^7 + A_{-3 }^\alpha A_{-1}^7
\}\tach{a}} ,\\
{\{A_{-1}^\mu  A_{-1}^8 A_{-1}^8 A_{-1}^8 + A_{-1}^\mu A_{-3}^8
\}\tach{a}} ,\\
{\{A_{-1}^\alpha A_{-1}^\beta A_{-1}^\gamma A_{-1}^\gamma + A_{-3
}^\alpha A_{-1}^\beta \}\tach{a}},\\
{\{A_{-1}^\alpha A_{-1}^7 A_{-1}^\beta A_{-1}^\beta - A_{-3 }^\alpha
A_{-1}^7 \}\tach{a}} ,\\
{\{A_{-1}^\alpha A_{-1}^\beta A_{-1}^7 A_{-1}^7 - A_{-3 }^\alpha
A_{-1}^\beta \}\tach{a}} ,\\
{\{A_{-1}^\alpha A_{-1}^\beta A_{-1}^\gamma A_{-1}^\delta - 2  A_{-3
}^\epsilon A_{-1}^\eta \}\tach{a}},\\
\settowidth{\mylength}{${}^{(2)}$}
{\{(A_{-1}^i A_{-1}^j + \delta^{ij} \frac{1}{3} A_{-1}^8 A_{-1}^8
)A_{-2}^l \}\tach{a}}{}^{(2)}\hspace*{-\mylength},\\
\settowidth{\mylength}{${}^{(2)}$}
{\{(A_{-1}^i A_{-1}^j + \delta^{ij} \frac{1}{3} A_{-1}^8 A_{-1}^8)
A_{-2}^- \}\tach{a}}{}^{(2)}\hspace*{-\mylength},\\
{\{A_{-1}^8 A_{-1}^8 A_{-2}^- + \frac{3}{2} \sum_{i=1}^8 A_{-1}^i
A_{-3}^i \}\tach{a}} ,\\
{\{A_{-1}^\alpha A_{-1}^\alpha A_{-1}^\alpha A_{-1}^\alpha    +  2
A_{-1}^\alpha A_{-3}^\alpha +3A_{-1}^7 A_{-3}^7 - \sum_{\rho=1}^{7}
A_{-1}^\rho A_{-3}^\rho \}\tach{a}} ,\\
{\{A_{-1}^\alpha A_{-1}^\alpha A_{-1}^7 A_{-1}^7 + 3 A_{-1}^\alpha
A_{-3}^\alpha + A_{-1}^7 A_{-3}^7  - \frac{2}{3} \sum_{\rho=1}^{7}
A_{-1}^\rho A_{-3}^\rho \}\tach{a}} ,\\
{\{A_{-1}^\mu A_{-1}^\nu A_{-1}^8 A_{-1}^8 - A_{-1}^\mu A_{-3}^\nu +
\delta^{\mu \nu} ( A_{-1}^8 A_{-3}^8 + \frac{2}{3} \sum_{\rho=1}^7
A_{-1}^\rho A_{-3}^\rho ) \}\tach{a}} ,\\
{\{A_{-1}^\mu A_{-1}^\nu A_{-1}^\sigma A_{-1}^8 + \frac{1}{3}
\delta^{\mu \nu} A_{-1}^8A_{-3}^\sigma + \frac{1}{3} \delta^{\nu
\sigma} A_{-1}^8 A_{-3}^\mu + \frac{1}{3}\delta^{\mu \sigma} A_{-1}^8
A_{-3}^\nu \}\tach{a}} ,\\
{\{A_{-1}^\alpha A_{-1}^\alpha A_{-1}^\beta A_{-1}^\beta -
A_{-1}^\alpha A_{-3}^\alpha - A_{-1}^\beta A_{-3}^\beta - A_{-1}^7
A_{-3}^7 +  \frac{1}{3} \sum_{\mu=1}^{7} A_{-1}^\mu A_{-3}^\mu
\}\tach{a}} .\\
%
%
\end{array}
\]
Here we use the following conventions: roman letters $(i,j,\ldots)$ run
from 1 to 8, greek letters from the middle of the alphabet
$(\mu,\nu,\ldots)$ run from 1 to 7 and greek letters from the
beginning of the alphabet $(\alpha,\beta,\ldots)$ run from 1 to 6,
with the exceptions ${}^{(1)} i \in \{7,8\}$ and ${}^{(2)}
(i,j)\neq(8,8)$

As can be seen, the calculated states fall into two classes, depending
on whether the indices of the occurring DDF operators are odd or
even. Recall that the DDF construction of level-2 states involves
intermediate states with momenta of the type $\frac12
m\delta$. Therefore precisely the operators with odd indices
generate states that do not lie on the lattice and it can be seen
that these are responsible for the more complicated basis elements.

In view of current work \cite{BaGeGuNi97} on the structure of the
space of missing states let us present a basis for the orthogonal
complement of $E_{10}^{(\bLambda_1)}$ within the full space of physical
states. A direct calculation using the string scalar product reveals
that the 53-dimensional orthogonal complement of $E_{10}^{(\bLambda_1)}$ in
$\fg^{(\bLambda_1)}_{I\!I_{9,1}}$ is spanned by the following states:
$$
\begin{array}{r}
A_{-1}^i A_{-3}^- \tach{a} ,\\
\textstyle\big\{A_{-1}^8 A_{-1}^8 A_{-2}^i - \frac13 \sum_{\mu=1}^7
A_{-1}^\mu A_{-1}^\mu A_{-2}^i \big\}\tach{a} ,\\
\big\{A_{-2}^-A_{-2}^- - 4 A_{-4}^-\big\}\tach{a} ,\\
\textstyle\big\{A_{-1}^\mu A_{-1}^8 A_{-1}^8 A_{-1}^8 - 3
\sum_{\nu=1}^7 A_{-1}^\nu A_{-1}^\nu A_{-1}^\mu A_{-1}^8
-2 A_{-1}^\mu A_{-3}^8 \hphantom{\big\}\tach{a},} \\
  +6 A_{-1}^8 A_{-3}^\mu\big\}\tach{a} ,\\
\textstyle\big\{A_{-1}^\alpha A_{-3}^7 +  A_{-1}^7 A_{-3}^\alpha - 2
A_{-1}^\alpha A_{-1}^7 A_{-1}^7 A_{-1}^7
-4 A_{-1}^\alpha A_{-1}^\alpha A_{-1}^\alpha
A_{-1}^7\hphantom{\big\}\tach{a},} \\
\textstyle + \frac32 \sum_{i=1}^8 A_{-1}^i A_{-1}^i A_{-1}^\alpha A_{-1}^7
\big\}\tach{a} ,\\
\textstyle\big\{A_{-3}^\alpha A_{-1}^\alpha -\frac{3}{2}A_{-3}^8
A_{-1}^8  + \frac{1}{2}A_{-3}^7 A_{-1}^7  - A_{-1}^\alpha
A_{-1}^\alpha A_{-1}^\alpha A_{-1}^\alpha\hphantom{\big\}\tach{a},}\\
\textstyle - \frac{1}{4}A_{-1}^7 A_{-1}^7 A_{-1}^7 A_{-1}^7
+\frac{3}{4}\sum_{i = 1}^{8} A_{-1}^i A_{-1}^i A_{-1}^\alpha
A_{-1}^\alpha\hphantom{\big\}\tach{a},} \\
\textstyle  + \frac{3}{8}\sum_{i= 1}^{8} A_{-1}^i A_{-1}^i A_{-1}^7
A_{-1}^7  - A_{-1}^\alpha A_{-1}^\alpha A_{-1}^7 A_{-1}^7
\hphantom{\big\}\tach{a},}\\
\textstyle + \frac{3}{8}\sum_{\mu = 1}^{7} A_{-1}^\mu A_{-1}^\mu
A_{-1}^8 A_{-1}^8  - \frac{3}{8}A_{-1}^8 A_{-1}^8 A_{-1}^8 A_{-1}^8
\big\} \tach{a},\\
\textstyle\big\{ A_{-1}^\alpha A_{-3}^\beta +  A_{-1}^\beta
 A_{-3}^\alpha +\frac12 A_{-1}^\alpha A_{-1}^\beta A_{-1}^\beta
 A_{-1}^\beta +\frac12 A_{-1}^\alpha A_{-1}^\alpha A_{-1}^\alpha
 A_{-1}^\beta
 \hphantom{\big\}\tach{a},} \\
\textstyle -\frac32 \sum_{\gamma=1 \atop \gamma \neq \alpha,\beta}^6
 A_{-1}^\alpha A_{-1}^\beta A_{-1}^\gamma A_{-1}^\gamma  + \frac32
 A_{-1}^\alpha A_{-1}^\beta A_{-1}^8 A_{-1}^8
 \hphantom{\big\}\tach{a},}\\
\textstyle + \frac32 A_{-1}^\alpha A_{-1}^\beta A_{-1}^7 A_{-1}^7
+ 4  A_{-1}^\gamma A_{-1}^\delta A_{-1}^\epsilon A_{-1}^\eta
\big\}\tach{a},\\
\textstyle\big\{\frac43 A_{-1}^8 A_{-3}^8 - \frac18 \sum_{\mu=1}^7
A_{-1}^\mu A_{-1}^\mu A_{-2}^-
+ \frac38 A_{-1}^8 A_{-1}^8 A_{-2}^-\hphantom{\big\}\tach{a}}\\
\textstyle \frac13 A_{-1}^8  A_{-1}^8 A_{-1}^8 A_{-1}^8 + \frac14
\sum_{\mu=1}^7 A_{-1}^\mu A_{-1}^\mu A_{-1}^8 A_{-1}^8
\big\}\tach{a},\\
\textstyle\big\{7 A_{-1}^8 A_{-3}^8 + \frac74 A_{-1}^8 A_{-1}^8
A_{-1}^8 A_{-1}^8 -\frac52 \sum_{\mu=1}^7 A_{-1}^ \mu A_{-1}^\mu
A_{-1}^8 A_{-1}^8
\hphantom{\big\}\tach{a},}\\
\textstyle -\frac14 \sum_{\mu,\nu=1}^7 A_{-1}^\mu A_{-1}^\mu
A_{-1}^\nu A_{-1}^\nu
 - \sum_{\mu=1}^7 A_{-1}^\mu A_{-3}^\mu \big\}\tach{a}.
\end{array}
$$
\section*{Acknowledgments}
We thank Hermann Nicolai and Kilian Koepsell for stimulating
discussions and Mark Turner for careful reading of the
manuscript. R.~W.~G. is very grateful to the Deutsche
Forschungsgemeinschaft for financial support enabling him to stay at
the Institute for Advanced Study.
\appendix
\section{Calculations}
The calculations where performed using a Lisp program and the symbolic
algebra system MAPLE.

We will first consider the decomposition of $\bLambda_1$ with $m=3$.
Since the calculations for the other decomposition are similar we will
treat this example in detail and skip the other case. One starts by
collecting all level-1 roots that upon commutation give states in this
particular root space. Since the DDF construction provides us with
explicit expressions for all level-1 roots this is easily done. We
arrive at the following list of commutators:
\begin{eqnarray}
\label{eq:comlist}
\eqalign{
\com{\tach{s}, A_{-3}^i \tach{r}}, &\com{A_{-1}^i \tach{s}, A_{-2}^j
\tach{r}},\\
\com{\tach{s}, A_{-2}^i A_{-1}^j \tach{r}}, \qquad& \com{A_{-1}^k
\tach{s},A_{-1}^i A_{-1}^j \tach{r}},\\
\com{\tach{s}, A_{-1}^i A_{-1}^j A_{-1}^k \tach{r}}.&}
\end{eqnarray}
where the indices $i,j,k$ run from 1 to 8. We will explain the
evaluation of commutators of this type by working out one example in
detail, for which we take the commutator $\com{\tach{s},
A_{-2}^{\alpha} A_{-1}^8\tach{r}}$.

The main technical problem for the calculations is that we have to
deal with two different objects. We started with the set of
oscillators $\{\alpha_m^\mu, \mu = 1 \ldots d, m \in \Z\}$ acting on
our Fock space $\cF$.  In terms of these we defined the vertex
operators and only here can we perform the explicit calculations. We
then restricted ourselves to the subspace of physical states $\cP^1$
and introduced the DDF operators providing us with a elegant way to
obtain explicit bases for the subspaces of $\cP^1$ of some definite
momentum. The price we have to pay for this elegance is that it is not
possible to perform explicit calculations using the DDF operators.

Returning to our example, we start by determining the corresponding
oscillator representations of the commuted states. We find
\begin{eqnarray}
\fl A_{-2}^{\alpha} A_{-1}^8 \tach{r}&=&\com{\bxi_\alpha(-1)\tach{\mbox{2}
\delta},\com{\bxi_8(-1)\tach{\delta},\tach{r}}}\nonumber\\
&=&\epsilon(\vk,\vr)\left\{\bxi_\alpha(-2)\bxi_8(-1)+
2\bxi_\alpha(-1)\bxi_8(-1)\bdelta(-1)\right\}\tach{r+\mbox{3}\bdelta},
\end{eqnarray}
where $\epsilon(\vk,\vr)$ denotes the cocycle factor due to the
twisted group algebra $\R\{I\!I_{9,1}\}$.
For the calculation of the commutator we need the vertex operator
corresponding to $\tach{r}$ which is given by the definition
(\ref{eq:vertexop})
\[
\cV(\tach{r},z)=\textstyle\exp(\int \vr_<(z) \d z) \e^{\vr} z^{\vr(0)}
\exp(\int \vr_>(z) \d z),
\]
Now we can calculate the commutator, for which we find the following
expression
\begin{eqnarray}
  \label{osz}
\fl\com{\tach{s},A^\alpha_{-2}A^8_{-1}\tach{r}}=\nonumber\\
\fl\qquad\epsilon\big\{ -\case{4}{3}\bxi_\alpha (-1)\bxi_8 (-3) -
\case12\bxi_8 (-2)\bxi_\alpha (-2) - \case{1}{3}\bxi_\alpha (-1)\bxi_8
(-1)^3\nonumber \\
\fl\qquad\phantom{\epsilon\big\{}
- \bxi_\alpha (-1)\bdelta(-1)\bxi_8 (-2) - \case12\bxi_8
(-1)\bdelta(-1)\bxi_\alpha (-2)\nonumber\\
\fl\qquad\phantom{\epsilon\big\{}
- \case12\bxi_8 (-1)\bxi_\alpha (-1)\bdelta(-1)^2 - \case12\bxi_8
(-2)\bLambda(-1)\bxi_\alpha (-1) - \case{4}{3}\sqrt{2}\bxi_\alpha
(-1)\bdelta(-3)\nonumber \\
\fl\qquad\phantom{\epsilon\big\{}
    + \case{1}{8}\sqrt{2}\bxi_\alpha (-2)\bLambda(-1)^2 +
\case{1}{24}\sqrt{2}\bxi_\alpha (-1)\bLambda(-1)^3 +
\case{3}{8}\bxi_\alpha (-2)\bdelta(-1)^2\sqrt{2}\nonumber\\
\fl\qquad\phantom{\epsilon\big\{}
- \case14\bxi_\alpha (-2)\sqrt{2}\bxi_8 (-1)^2+
\case{7}{12}\bxi_\alpha (-1)\bdelta(-1)^3\sqrt{2} \nonumber \\
\fl\qquad\phantom{\epsilon\big\{}
+ \case14\sqrt{2}\bxi_\alpha (-2)\bLambda(-2) -
\case{3}{4}\sqrt{2}\bxi_\alpha (-2)\bdelta(-2) +
\case{1}{3}\sqrt{2}\bxi_\alpha (-1)\bLambda(-3)\nonumber \\
\fl\qquad\phantom{\epsilon\big\{}
- \bxi_8 (-1)\bxi_\alpha (-1)\bdelta(-2) - \bxi_\alpha
(-1)\sqrt{2}\bxi_8 (-1)\bxi_8 (-2)\nonumber\\
\fl\qquad\phantom{\epsilon\big\{}
- \case12\bxi_\alpha (-1)\bdelta(-1)\sqrt{2}\bxi_8 (-1)^2 -
\case14\bxi_\alpha (-1)\bLambda(-1)\sqrt{2}\bxi_8 (-1)^2\nonumber \\
\fl\qquad\phantom{\epsilon\big\{}
- \case12\bxi_\alpha (-2)\bdelta(-1)\bLambda(-1)\sqrt{2} -
\case{5}{8}\bxi_\alpha
(-1)\bdelta(-1)^2\bLambda(-1)\sqrt{2}\nonumber\\
\fl\qquad\phantom{\epsilon\big\{}
+ \case14\sqrt{2}\bxi_\alpha (-1)\bLambda(-2)\bLambda(-1) -
\case{3}{4}\sqrt{2}\bxi_\alpha (-1)\bLambda(-1)\bdelta(-2)\nonumber \\
\fl\qquad\phantom{\epsilon\big\{}
- \case12\bxi_\alpha (-1)\bxi_8 (-1)\bdelta(-1)\bLambda(-1) -
\case12\sqrt{2}\bxi_\alpha (-1)\bdelta(-1)\mb{\delta}(-2) \big\}
\tach{a},
\end{eqnarray}
where $\epsilon \equiv \epsilon(\vs-\vk,\vr)$.  To compare this result
with the basis of the subspace of physical states we have to reexpress
the oscillators in terms of DDF operators.  Here we encounter another
subtlety: we used the general Fock space for calculations where
elements of the Lie algebra of physical states are defined only modulo
null states, that is, states of the form $\rL{-1}\psi$ for a general
state $\psi$. To obtain equivalence and to explicitly solve the system
of linear equations we have to include a general null state. We arrive
at
\begin{eqnarray}
\label{_a8}
\lefteqn{
\com{\tach{s}, A_{-2}^{\alpha} A_{-1}^8\tach{r}} =} \nonumber\\
& & \epsilon \big\{
-\case{1}{2}\sqrt{2}A_{-4}^\alpha  - \case{1}{2}\sqrt{2}A_{-1}^\alpha
A_{-1}^8 A_{-2}^8  + \case{1}{3}A_{-1}^\alpha A_{-1}^8 A_{-1}^8
 A_{-1}^8  + \case{1}{3}A_{-1}^\alpha A_{-3}^8 \big\} \tach{a}
 \nonumber \\
& &
+ \rL{-1} \big\{ \case{5}{8}\bxi_\alpha
(-1)\bdelta(-1)^2\sqrt{2} - \case{1}{24}
\bxi_\alpha (-1)\bLambda(-1)^2\sqrt{2} + \case{1}{4}\bxi_\alpha (-1
)\sqrt{2}\bxi_8 (-1)^2 \nonumber\\
& & \phantom{+ \rL{-1} \big\{}
-\case{1}{2}\bxi_8 (-2)\bxi_\alpha (-1)
-\case{1}{12}\bxi_\alpha (-2)\bLambda(-1)\sqrt{2} + \case{1}{
2}\bdelta(-1)\bxi_\alpha (-2)\sqrt{2} + \case{1}{6}\bxi_\alpha
(-3)\sqrt{2}\nonumber \\
& & \phantom{+ \rL{-1} \big\{}
+ \case{3}{4}\bxi_\alpha (-1)\sqrt{2}- \case{1}{6}\bxi_\alpha
 (-1)\sqrt{2}\bLambda(-2) \bdelta(-2)
   - \case{1}{2}\bxi_8 (-1)\bdelta(-1)\bxi_\alpha (-1)
\big\} \tach{\mbox{\unboldmath$\bLambda_1$\boldmath}}.
\end{eqnarray}
Without calculation we record another commutator
\begin{eqnarray}
\label{8_a}
\lefteqn{
\com{A_{-1}^8 \tach{s}, A_{-2}^{\alpha} \tach{r}}=}  \nonumber\\
& &\epsilon \big\{ \case{1}{2}\sqrt{2}A_{-4}^\alpha  -
\case{1}{2}\sqrt{2}A_{-1}^\alpha A_{-1}^8 A_{-2}^8  -
\case{1}{3}A_{-1}^\alpha A_{-3}^8  - \case{1}{3}A_{-1}^\alpha A_{-1}^8
A_{-1}^8 A_{-1}^8 \big\} \tach{a},
\end{eqnarray}
where we suppressed the $\rL{-1}\big\{ \ldots \big\}$-term, as we will
do in all other calculations. To simplify these elements of $E_{10}$
we take suitable linear combinations. We immediately get
\[
-\case1{2\epsilon}\sqrt{2}[(\ref{8_a})+(\ref{_a8})]=A_{-1}^\alpha
A_{-1}^8 A_{-2}^8
\]
as our first element of a basis for $E_{10}^{(\bLambda_1)}$. Another
possibility is acting on equations like (\ref{_a8}) with the little
Weyl group.  Recalling that $w_0$ leaves $\bxi_1,\ldots \bxi_7$
invariant and changes the sign of $\bxi_8$ we have:
\[
\case1{2\epsilon}\sqrt{2}[(\ref{8_a})-w_0(\ref{_a8})]=A_{-4}^\alpha.
\]
Evaluating all commutators and simplifying the resulting equation
leads to a basis for a 516-dimensional subspace:
\[\fl
\begin{array}{r}
\settowidth{\mylength}{${}^{(1)}$}
A_{-1}^i A_{-1}^j A_{-2} ^l \tach{a}^{(1)}\hspace*{-\mylength},\\
A_{-2}^i A_{-2}^j\tach{a} ,\\
A_{-2}^i A_{-2}^-\tach{a} ,\\
A_{-4}^i\tach{a}  ,\\
\{A_{-1}^8 A_{-1}^8 A_{-1}^8 A_{-1}^8 - 2 A_{-1}^8 A_{-3}^8\}\tach{a} ,\\
\{A_{-1}^\mu  A_{-1}^8 A_{-1}^8 A_{-1}^8 + A_{-1}^\mu A_{-3}^8
\}\tach{a}  ,\\
\{A_{-1}^\mu A_{-3}^\nu - A_{-3}^\mu A_{-1}^\nu\}\tach{a}  ,\\
\{3 A_{-1}^\mu A_{-3}^8 +  A_{-1}^8 A_{-3}^\mu\}\tach{a}  ,\\
\{A_{-1}^\mu A_{-1}^8 A_{-2}^-\}\tach{a}  ,\\
\{A_{-1}^\mu A_{-1}^\mu A_{-1}^i + \case{1}{3} A_{-1}^8 A_{-1}^8
A_{-2}^i\}\tach{a}  ,\\
\settowidth{\mylength}{$\}\tach{a},$}
\{A_{-1}^\beta A_{-3}^\alpha - A_{-1}^\alpha A_{-1}^\beta A_{-1}^8
A_{-1}^8 \hspace*{\mylength} \\
+ \delta^{\alpha \beta} ( \case{1}{5} A_{-1}^8 A_{-1}^8 A_{-2}^- +
\case{1}{10} \sum_{\mu=1}^7
A_{-1}^\mu A_{-1}^\mu A_{-1}^8 A_{-1}^8 )\}\tach{a} ,\\
\settowidth{\mylength}{$\}\tach{a},$}
\{A_{-1}^\alpha A_{-1}^\beta A_{-1}^- + \case{1}{2} A_{-1}^\alpha
A_{-1}^\beta A_{-1}^8  A_{-1}^8 + \frac{1}{2} \sum_{\mu=1}^7
A_{-1}^\mu A_{-1}^\mu A_{-1}^\alpha A_{-1}^\beta \hspace*{\mylength}
\\
+ \delta^{\alpha \beta} ( A_{-1}^8 A_{-3}^8 + \frac{1}{10}
\sum_{\mu=1}^7 A_{-1}^\mu A_{-1}^\mu A_{-1}^8 A_{-1}^8 +\frac{1}{5}
A_{-1}^8 A_{-1}^8 A_{-2}^-)\}\tach{a}  ,\\
\{A_{-1}^\alpha A_{-1}^\beta A_{-1}^\gamma A_{-1}^8
+ \frac{1}{3} \delta^{\alpha \beta} A_{-1}^8 A_{-3}^\gamma
+ \frac{1}{3} \delta^{\beta \gamma} A_{-1}^8 A_{-3}^\alpha +
\frac{1}{3} \delta^{\alpha \gamma} A_{-1}^8 A_{-3}^\beta \}\tach{a}
,\\
\{A_{-2}^- A_{-2}^- - A_{-1}^8 A_{-1}^8 A_{-2}^- - \frac{1}{4} \sum_{
\mu , \nu = 1}^{7} A_{-1}^\mu A_{-1}^\mu A_{-1}^\nu A_{-1}^\nu -
\frac{3}{2} A_{-1}^8 A_{-3}^8 \}\tach{a}.
\end{array}
\]
We use the following conventions: roman letters $(i,j,\ldots)$ run
from 1 to 8  and greek letters from the
middle of the alphabet $(\mu,\nu,\ldots)$ run from 1 to 7, and we have
the exception ${}^{(1)} i \neq j$.

For the decomposition with $m=2$ we have to evaluate the commutators
\begin{equation}
\eqalign{
\com{\tach{s'}, A_{-1}^i A_{-1}^j \tach{r'}}\\
 \com{A_{-1}^i\tach{s'},A_{-1}^j \tach{r'}}}
\end{equation}
Analogous calculations here and combination of the two results lead to
the basis given in the main part.

\end{document}